\documentclass[aps,prd,superscriptaddress]{revtex4}

\usepackage{amsmath}
\usepackage{latexsym}
\usepackage{epsfig}
\usepackage{graphicx}
\usepackage{psfrag}
\usepackage{subfigure}
\usepackage{placeins} 

\usepackage{times,mathptmx}     
\usepackage{amsfonts}		
\usepackage{amsmath}		
\usepackage{amssymb}		
\usepackage{upgreek}		
\usepackage{wasysym}
\usepackage{multirow}
\usepackage[english]{babel}
\usepackage{amsmath}
\usepackage{units}

\newcommand{\td}{\text{d}}

\newcommand*{\etal}{\emph{et al.}}

\newcommand*{\ie}{\emph{i.e.}}

\newcommand{\bea}{\begin{eqnarray}}
\newcommand{\eea}{\end{eqnarray}}
\newcommand{\be}{\begin{equation}}
\newcommand{\ee}{\end{equation}}

\def\gtsim{\lower-0.45ex\hbox{$>$}\kern-0.77em\lower0.55ex\hbox{$\sim$}}
\def\ltsim{\lower-0.45ex\hbox{$<$}\kern-0.77em\lower0.55ex\hbox{$\sim$}}

\begin{document}



\title{QCD Evolution of Jets in the Quark-Gluon Plasma}



\author{S. Domdey}
\email[]{Domdey@tphys.uni-heidelberg.de}
\affiliation{Institut f\"ur Theoretische Physik, Universit\"at Heidelberg, Germany}

\author{G. Ingelman}
\email[]{Gunnar.Ingelman@tsl.uu.se}
\affiliation{High Energy Physics, Uppsala University, Box 535, S-75121 Uppsala, Sweden }

\author{H.J. Pirner}
\email[]{pir@tphys.uni-heidelberg.de}
\affiliation{Institut f\"ur Theoretische Physik, Universit\"at Heidelberg, Germany}

\author{J. Rathsman}
\email[]{Johan.Rathsman@tsl.uu.se}
\affiliation{High Energy Physics, Uppsala University, Box 535, S-75121 Uppsala, Sweden }

\author{J. Stachel }
\email[]{johanna.stachel@physik.uni-heidelberg.de}
\affiliation{Physikalisches Institut, Universit\"at Heidelberg, Philosophenweg 12, DE-69120 Heidelberg, Germany}

\author{K. Zapp}
\email[]{zapp@pi0.physi.uni-heidelberg.de}
\affiliation{Physikalisches Institut, Universit\"at Heidelberg, Philosophenweg 12, DE-69120 Heidelberg, Germany}


\date{\today}

\begin{abstract}

 The quark-gluon plasma (QGP) can be explored in
  relativistic heavy ion collisions by the jet quenching signature,
  i.e. by the energy loss of a high energy quark or gluon traversing
  the plasma.  We introduce a novel QCD evolution formalism in the
  leading logarithm approximation, where normal parton radiation is
  interleaved with scattering on the plasma gluons occuring at
  a similar time scale.  The idea is elaborated in two approaches. One extends the DGLAP evolution equations for fragmentation functions to include scatterings in the medium, which facilitates numerical solutions for comparison with data and provides a basis for a Monte Carlo implementation.
  The other approach is more general by including also the transverse momentum dependence of the jet evolution, which allows a separation of the scales also for the scattering term and provides a basis for analytical investigations. The two approaches are shown to be related and give the same  characteristic softening of the jet depending on the temperature of the plasma. A substantial effect is found at the RHIC energy and is further enhanced at LHC. Systematic studies of data on the energy loss could, therefore, demonstrate the existence of the QGP and probe its properties.
\end{abstract}




\maketitle


\section{Introduction}\label{sec-intro}
The quark-gluon plasma (QGP) is expected to be formed in high energy
collisions of nuclei when the energy density is large enough. As the
plasma expands it cools down and undergoes a phase transition to
ordinary hadrons. Different observables have been suggested to provide
a signal for the existence of the plasma. One of the most promising
signals, that can also reveal detailed information on the properties
of the QGP, appears at present to be jet quenching, i.e. the energy loss
of an energetic quark or gluon traversing the plasma resulting in a
jet of hadrons, or single hadrons, with reduced energy \cite{Bjorken:1982tu,Wang:1991xy,Baier:1996kr}. Thus, one may consider a normal hard QCD parton-parton scattering event to 
take place in the environment of the QGP which has been produced in the nucleus-nucleus collision. The QGP acts as a
background with which the hard-scattered partons may interact.

Such jet quenching phenomena have indeed been observed at RHIC \cite{Arsene:2004fa,Back:2004je,Adams:2005dq,Adcox:2004mh,whitepapers}, but it
remains to be demonstrated that they are really caused by a QGP. 
This requires a better understanding of the QCD processes involved and this paper contributes to this. Once a proper framework has been established, one can use the modified jet properties as a trustworthy signal to claim discovery of the QGP, and also as a tool to study its properties.

Until recently, theoretical efforts have concentrated on medium-induced gluon bremsstrahlung as the dominant energy loss mechanism 
\cite{wang, gyulassy, wiedemann, Gyulassy:2003mc,Armesto:2007dt,Wang:2006qr}. 
However, collisional energy loss caused by multiple scatterings 
in the medium has also been considered \cite{Lokhtin:1998ya,Zapp:2005kt,Adil:2006ei} and 
demonstrated \cite{Zapp:2005kt} to contribute significantly to the jet
quenching effect observed at RHIC.

Although these scatterings occur over the few-fermi extended plasma,
they overlap in time with normal perturbative QCD (pQCD) gluon
radiation (the parton cascade) which is time dilated in the QGP rest frame. Therefore, such radiation and interactions with the plasma must be treated in a common framework. This is the objective of this paper.

The energy available at RHIC limits the transverse momentum of the
hard-scattered parton such that one mainly observes single high $p_\perp$
hadrons, rather than complete jets. With heavy ion collisions in LHC,
however, the energy is substantially larger and should give access to
the study of jets and how their properties are affected by the
underlying interaction with the plasma. The larger momentum transfer
should also establish a safer ground for a perturbative QCD treatment.

In Section \ref{sec-basics}, we introduce scattering with the plasma into the
familiar jet evolution DGLAP equation \cite{Gribov:1972ri,Altarelli:1977zs,Dokshitzer:1977sg}. In this way the scattering can be solved numerically without further approximations and this evolution equation provides a basis for development of a modified parton shower Monte Carlo algorithm \cite{Zapp-MC}.
In Section \ref{sec-transverse}, we start from a more general evolution equation taking also the transverse momentum into account and introduce the scattering. We show how this modifies the evolution for the dominant small momentum transfer scatterings. Using appropriate approximations, we obtain a medium-modified evolution equation which is suitable for analytical investigations. We also show that by integrating out the transverse momentum dependence the medium-modified DGLAP equations of Section \ref{sec-basics} is recovered. In Section \ref{sec-numerical}, both approaches are used to obtain numerical results on how the fragmentation functions are affected by this jet quenching mechanism. Section \ref{sec-conclusions} concludes with a discussion of our results.

\section{Basics of jet evolution in the quark-gluon plasma}\label{sec-basics}
A quark or gluon that emerges from a hard scattering can be virtual ($p^2\gg m^2$) and therefore it will radiate successively to reduce its virtuality and become on mass-shell. This leads to a parton shower and scaling
violations in the jet fragmentation functions as described by the \textsc{DGLAP} equations \cite{Gribov:1972ri,Altarelli:1977zs,Dokshitzer:1977sg} and other more sophisticated evolution equations \cite {Bassetto:1979nt}.

The formation of this parton shower does not happen instantaneously, but needs a certain time. In the lab frame, i.e.\ in the rest frame of the nucleus-nucleus collision and the plasma, the lifetime of a virtual parton can be estimated from the uncertainty principle to be $E/Q^2$, where $E$ is the parton's energy and $Q$ its virtual mass. In the case of a parton shower, however, the relevant time is the time a virtual state needs to evolve in virtuality from $Q$ to $Q+d Q$,
\begin{equation}\label{dtau}
d\tau= \frac{E}{Q^2}\frac{dQ^2}{Q^2}
\end{equation}
The time a parton needs to reduce its virtuality from the starting
scale $Q_\text{i} \simeq E \simeq 100\mbox{ GeV}$ to the hadronisation
scale $Q_0 \simeq 1$ GeV can then be
estimated as $\tau = \int d\tau \approx \frac{E}{Q_0^2}-\frac{1}{E}$. Thus, even though a high energy parton with large virtuality will reduce its virtuality rapidly, the overall lifetime can be considerable and of the order $E/(1 GeV)^2$ or several fermi. The formation time $\tau_i$ of the quark-gluon plasma  \cite{Adcox:2004mh} is estimated to be of the order 0.2 fm at \textsc{Rhic} and likely less than 0.1 fm at \textsc{Lhc}, but the plasma lifetime $\tau_{Plasma}\simeq\tau_i(T_i/T_c)^3$ may be long due to the cooling from the initial temperature to the critical temperature. Taking into account these two time scales of the parton and the
surrounding medium we arrive at the conclusion that the parton shower evolution overlaps in time with the plasma phase. Therefore, energy loss through radiation and scattering has to be treated together in a common framework.

In this paper we present a study of a parton cascade that includes scattering off the partons in the QGP as a modification to the well known QCD evolution in vacuum. This is achieved by adding a scattering term in the DGLAP equations.  Modified DGLAP evolution equations due to medium-induced radiation has been considered in \cite{Armesto:2007dt} for the QGP and for deep inelastic scattering off nuclei in \cite{Wang:2001ifa}. The literature on jet quenching in the QGP also deals with effects like energy loss of on-shell and off-shell partons and modified hadronisation at $Q_0$ \cite{wang, gyulassy, wiedemann}. Our approach differs from these by explicitly handling off-shell partons and the evolution of the shower in the presence of the medium.

We assume here that the hadronisation at the infrared cut-off scale $Q_0$ is unchanged and use standard vacuum fragmentation functions, since the plasma is essentially gone by the time the parton has evolved down to this low virtuality. Furthermore, we are here primarily interested in the high-$p_\perp$ particles that stand out from the underlying background of hadrons from the hadronisation of the plasma and such higher energy hadrons have a longer formation time in the lab frame due to their larger Lorentz $\gamma$ factor. 
In this and the next section we concentrate on developing the
formalism for the evolution at the parton level and thereafter add on
hadronisation through a well established parametrisation.

Let us consider a hard process where a parton with high virtuality and high energy is produced. Due to parton splitting and parton scattering a parton
shower emerges from this source. We define the fragmentation function
$D_i^j(x,Q^2)$ as the probability density for the hard parton $i$ with virtuality $Q^2$ to fragment into a parton or hadron $j$ which takes a fraction $x$ of the initial parton energy.
In this first study we restrict our parton level treatment to a purely gluonic system, which should give the dominating behaviour and most essential results. The indices on the fragmentation functions can then be dropped and the formalism becomes simplified also in other aspects.

In the shower evolution, an arbitrary intermediate gluon with a fraction $y$ of the original energy $E$ loses a fraction $z$ through radiation or scattering, and becomes an outgoing gluon with energy $z(yE)$.
The parton cascade occurs through the normal splitting process $g \to g+g$ with differential probability
\begin{equation}
 \text{d}\mathcal{P}_\text{splitting }\left(z,Q^2\right) = \frac{\alpha_s(Q^2)}{2\pi} \hat P(z)\text{d}z \frac{\text{d}Q^2}{Q^2}
\end{equation}
which is of first order in $\alpha_s$ and written in terms of the unregularised splitting function 
\begin{equation}
 \hat P (z) =C_A\left[ \frac{z}{(1-z)} + \frac{1-z}{z}+z(1-z)\right]
\end{equation}
The scattering process $g+g \to g+g$ is similarly given by the differential probability 
\begin{equation}\label{P-scat}
 \text{d}\mathcal{P}_\text{scat}\left(z,yE,Q^2\right) = 
\frac{d\sigma}{dz} n_g(T) dz d\tau =
 \alpha_s^2(Q^2) \hat K\left(z,yE,Q^2\right) \text{d}z\frac{\text{d}Q^2}{Q^2}
\end{equation}
as the product of the scattering cross-section, the density of target gluons in the plasma and the time (or equivalently distance) the gluon traverses the plasma. We introduce the scattering function $\hat K$ in analogy with the splitting function to combine gluon splittings and gluon scatterings in a generalized DGLAP equation for the QCD evolution of a jet in the plasma medium 
\begin{eqnarray}\label{Sk}
\frac{\partial\, D(x,Q^2)}{\partial\, \ln Q^2} 
  &=&   \int \limits_x^1\!\! \frac{\td z}{z}\, \left\{
\frac{\alpha_s(Q^2)}{2\pi} 2 \hat P(z)
D(\frac{x}{z},Q^2) + \alpha_s^2(Q^2) \hat K\left(z,\frac{x}{z}E,Q^2\right)
D(\frac{x}{z},Q^2)\right\}  \nonumber \\
  & & \qquad - \int \limits_0^1\!\! \td z \, \left\{ \frac{\alpha_s(Q^2)}{2\pi}
\hat P(z) D(x,Q^2) + \alpha_s^2(Q^2) \hat K(z,xE,Q^2) D(x,Q^2)\right\}
\end{eqnarray}
where in the first line $\frac{x}{z}=y$, i.e.\ the intermediate energy fraction $y$  is integrated out via $\int dy \ldots\delta(y-x/z)$.
The negative contribution in the second line is present when using the unregularized splitting function instead of the regularized one with the 'plus prescription' \cite{Ellis}. This formulation gives a clear physical interpretation, namely that the first line in Eq. (\ref{Sk}) gives the {\em gain} of a gluon at $x$ which was at $yE$ before the splitting or scattering, and the second line gives the {\em loss} of a gluon with energy $xE$ before the splitting or scattering. We also note that the two loss terms can be thought of as ensuring that probability is conserved. For the splitting loss term it leads after integration of the DGLAP equation to the so-called Sudakov factor, whereas from the scattering loss term one gets a factor $\sim \exp{\left( -\int \frac{d\sigma}{dz} n_g\, dz d\tau \right)}$ corresponding to the attenuation of a beam.

To obtain the scattering function $\hat K$, we develop the different parts of Eq. (\ref{P-scat}). For the $g+g\to g+g$ scattering cross section we use the leading order pQCD formula
\begin{equation}\label{eq:dsigmadt}
\frac{d\sigma}{d|t|} = \alpha_\text{s}^2(Q^2)\frac{9}{4}\,
\frac{2\pi}{(|t| + \mu_\text{D}^2)^2}
\end{equation}
which has been regularized  by the Debye mass $\mu_\text{D}$ of the plasma 
gluons which screens the interaction such that this expression can be applied  
down to $|t|=0$. Furthermore, the scale in $\alpha_\text{s}$ is chosen as the 
gluon virtuality $Q^2$ and is thereby in the perturbative region above 
$Q^2_0\sim 1$ GeV$^2$. This is as for the usual DGLAP parton splitting and 
describes the weaker interaction of a more virtual parton having a smaller 
gluon cloud. We use the one loop approximation
\begin{equation}
\alpha_s(Q^2)= \frac{1}{b \log(Q^2/\Lambda^2)}\qquad \mbox{with} \qquad
b =\frac{11}{4 \pi}.
\end{equation}
for the running strong coupling with $\Lambda=0.25$ GeV. The cross section in Eq.\ (\ref{eq:dsigmadt}) implicitly assumes that the virtual mass of the fast parton does not change in the scattering, and applies for a fixed centre-of-mass energy, i.e.\ fixed $yE$. For a scattering centre at rest with mass $m_\text{s}$, the squared momentum transfer is then 
\begin{equation}
 t = 2 m_\text{s} yE (z-1)
 \label{eq:momtransf}
\end{equation}

The scattering is on gluons in the relativistic plasma of temperature $T$ with density
\begin{equation}\label{eq:density}
 n_g(T)=\frac{16}{\pi^2} \zeta(3) T^3
\end{equation}
where $\zeta$ is the Riemann Zeta function with $\zeta(3)\simeq 1.20$. We neglect the motion of the scattering centers, which is a good approximation at large jet energies, i.e.\ $xE \gg T$. For small $x$ this is not necessarily the case, but in this region the DGLAP equations should also be improved by soft gluon resummation for an accurate description of the shower evolution. The Debye mass $\mu_\text{D}\propto T$ \cite{Braun:2006vd} and the mass of the scattering centre is taken as $m_\text{s}=\mu_\text{D}/\sqrt{2}$.

Including also $d\tau$ from Eq.\ (\ref{dtau}) we obtain the scattering function as a function of the gluon energy ${\cal E}$,
\begin{equation}\label{K-hat}
\hat K\left(z,{\cal E},Q^2\right) = \frac{9 \pi\ n_g(T) m_\text{s}{\cal E}^2}
{Q^2 (2m_\text{s} {\cal E}(1-z) + \mu_\text{D}^2)^2}
\end{equation}
where ${\cal E}=yE=\frac{x}{z}E$ in the gain term of Eq.\ (\ref{Sk}) and 
${\cal E}=xE$ in the loss term. 

This $\hat{K}$ contains the essential dynamics of the scattering, in analogy 
with the splitting function $\hat{P}$ in Eq.\ (\ref{Sk}).
For large momentum transfers, i.e.\ 
$-t=2m_\text{s} {\cal E} (1-z)\gg \mu_\text{D}^2$, 
the scattering function is suppressed as   $\hat K\sim 1/(Q^2(1-z)^2)$ 
 as a higher twist process.
For small momentum transfers,  on the other hand, the scattering function 
$\hat K\sim {\cal E}^2 /Q^2$.
Thus, $\hat{K}$ may compensate for the additional factor $\alpha_s$ in the 
scattering term as compared to the splitting term in the evolution 
Eq.\ (\ref{Sk})  for large ${\cal E}$ and/or large $x$.

The jet quenching effect is then given by the difference between the
scattering gain and loss terms in Eq.\ (\ref{Sk}),  i.e.\ 
\begin{eqnarray} \label{Sk2}
\hat{S}(x,Q^2) 
  &=&   \int \limits_x^1\!\! \frac{\td z}{z}\, 
 \alpha_s^2(Q^2) \hat K\left(z,\frac{x}{z}E,Q^2\right)
D(\frac{x}{z},Q^2) 
    - \int \limits_0^1\!\! \td z \,  \alpha_s^2(Q^2) \hat K(z,xE,Q^2) D(x,Q^2)
\end{eqnarray}
For leading particles in the jet,  i.e.\ for $x$ close to 1,
 and for large energies, $x(1-x)E \gg m_\text{s}$,
the gain term in Eq.\ (\ref{Sk2}) can be Taylor expanded  
(with $1/z=1+(1-z)+\ldots $ such that 
$D(x/z,Q^2)=D(x,Q^2)+x(1-z)\frac { \partial D(x,Q^2)}{\partial x} +\ldots $
and $K\left(z,\frac{x}{z}E,Q^2\right)= K\left(z,xE,Q^2\right)+\ldots$) giving
\begin{eqnarray} \label{Sk3}
\hat{S}(x,Q^2) 
  &=&   \int \limits_x^1\!\! \td z \, 
 \alpha_s^2(Q^2) \hat K\left(z,{x}E,Q^2\right)(1-z)
\left( D(x,Q^2)+x \frac { \partial D(x,Q^2)}{\partial x}\right) \nonumber \\
&&
    - \int \limits_0^x\!\! \td z \,  \alpha_s^2(Q^2) \hat K(z,xE,Q^2) D(x,Q^2).
\end{eqnarray}
Performing the integral over $z$ then gives
\begin{eqnarray} \label{Sk4}
\hat{S}(x \simeq 1,Q^2) 
  & = & 
\frac{\alpha_s^2(Q^2) 9 \pi n_g(T)}{4m_\text{s}Q^2} 
\left[\left( D(x,Q^2)+x \frac { \partial D(x,Q^2)}{\partial x}\right)  
\ln\frac{m_\text{s}+x(1-x)E}{m_\text{s}} + f(x) \right] \quad .
\end{eqnarray}
where $f(x)$ is independent of the energy for $x(1-x)E \gg m_\text{s}$.
Thus, for large energies and large $x$ such that, 
$m_\text{s}/E \ll 1-x \ll 1$, the effects of the 
scattering will grow logarithmically with the energy as a result of the 
integration limit $z=x$ corresponding to the maximum
momentum transfer squared $|t|=2m_\text{s}Ex(1-x)$ in the scattering gain term.

\section{Transverse broadening in jet evolution}\label{sec-transverse}
The parton transverse momenta can be explicitly taken into account in the jet evolution. Such an evolution equation in the vacuum was first derived in leading logarithmic approximation in ref. \cite{Bassetto:1979nt}. Recently two papers \cite{Ceccopieri:2005zz,Ceccopieri:2007ek} have appeared which apply the same framework of transverse-momentum-dependent QCD evolution equations to deep inelastic scattering. Here we generalize this to the case of jet evolution in the quark-gluon plasma by introducing scatterings as in the previous section. 

Transverse momentum of the tagged parton is an important tool to diagnose the history of the parton in the medium. In deep inelastic scattering in nuclei one can trace parton scattering and from the mean $p_\perp^2$ read off the length of the trajectory before hadron formation. In heavy ion scattering the jet cone related to the transverse momenta of partons will also reflect the scattering history. 

As before we restrict ourselves to gluon splitting and gluon scattering, 
since gluon dynamics dominates these processes. The ``fragmentation function'' 
$D(x,Q^2,\vec p_\perp)$ gives the probability for an initial gluon to convert 
into a gluon with momentum fraction $x$, virtuality $Q^2$ and transverse 
momentum $\vec p_\perp$ in the course of splittings and scatterings which 
give the gluon $\vec q_\perp$ kicks. The evolution equation for this multiple 
differential function has the following form, when we generalize it to include 
scattering in the medium 
\begin{eqnarray}\label{pt-evolution}
&&\frac{ \partial\, D(x,Q^2,\vec p_\perp)}{\partial\, \ln{Q^2}} =\nonumber\\ 
&&\frac{\alpha_s(Q^2)}{2 \pi}
\int_{x}^1
\frac{dz}{z} P(z,\alpha_s(Q^2))\frac{ d^2\vec q_\perp}{\pi} 
\delta\left(z(1-z)Q^2-\frac{Q_0^2}{4}-q_\perp^2\right) 
D\left(\frac{x}{z},Q^2,\vec p_\perp-\frac{x}{z} \vec q_\perp\right)+S(x,Q^2,\vec p_\perp)\nonumber\\
\end{eqnarray}
Both gains and losses from the evolution are here included by the use of the regularized splitting function $P(z)$ (instead of the unregularized $\hat P(z)$ in Eq.~(\ref{Sk})). The first, integral term is the normal evolution in the vacuum. This accounts for the mass constraint $z(1-z)Q^2=Q_0^2/4+q_\perp^2$ arising in the splitting with momentum fractions $z$ and $1-z$. The transverse momenta appear together with longitudinal momentum fractions to guarantee boost invariance. 
After integration over transverse momentum one obtains the standard DGLAP equation. 

The second term $S(x,Q^2,\vec{p}_\perp)$ accounts for the scattering on gluons in the plasma with temperature $T$ and the gluon density $n_g(T)$ in Eq. 
(\ref{eq:density}). The scatterings change the transverse momentum of the leading fast parton by giving $\vec q_\perp$ kicks, but are here assumed to not change the mass scale or virtuality of the fast parton. This is strictly only true for small momentum transfers, i.e.\ small angle scattering.
The time scale for interaction with the plasma $d\tau= \frac{E}{Q^2} \frac{dQ^2}{Q^2}$ (Eq.\ \ref{dtau}) depends on the virtuality. As before, this gives the connection between the $Q^2$-evolution time scale and the scattering time scale and shows that the parton cascade is interleaved with scattering in the plasma. Combining these ingredients we obtain 
\begin{eqnarray}\label{Sder_full}
S(x, Q^2,\vec p_\perp)&=& \frac{ E}{Q^2} n_g(T) \int _x^1 dy \int d^2\vec q_\perp 
\frac{d\sigma}{d^2\vec q_\perp}\times\nonumber\\
&&\times \left(y D(y,Q^2,\vec p_\perp-y \vec q_\perp)-x D(x,Q^2,\vec p_\perp)\right)
\delta\left(y- x-\frac{q_\perp^2}{2 m_\text{s} E}\right)
\end{eqnarray}
with a gain term for scattering into the considered $p_\perp$ bin and a loss term for scattering out of it. Both are weighted with the parton-parton differential cross section. 
The scattering partners of the fast parton are ``quasifree'' gluons with mass
$m_\text{s}$ in the plasma. Scattering  makes the fast parton lose
energy which is absorbed as recoil energy  $q_\perp^2/(2m_\text{s})$ by the plasma
parton.
In contrast to the previous section we here restrict to allow only soft 
scatterings, i.e.\  $q_\perp^2 \sim \mu_D^2$, such that 
$\Delta x= y-x = q_\perp^2/(2 m_\text{s} E)$ is small.

In this paper we restrict ourselves to the $p_\perp$-integrated parton fragmentation function 
$D (x,Q^2)=\int  d^2\vec p_\perp D (x,Q^2,\vec p_\perp)$.
The resulting $p_\perp$-integrated equation for splitting and scattering is then
\begin{equation}
\frac{ \partial\, D(x,Q^2)}{\partial\, \ln{Q^2}} =\frac{\alpha_s(Q^2)}{2 \pi} \int_{x}^1 \frac{dz}{z} P(z,\alpha_s(Q^2)) 
D\left(\frac{x}{z},Q^2\right)+\hat S(x,Q^2)\nonumber\\
\end{equation}
where 
\begin{eqnarray}\label{S_der}
\hat{S}(x,Q^2)=\frac{ n_g(T) \sigma \langle q_\perp^2\rangle}{ 2 m_\text{s} Q^2} 
 \left(D(x, Q^2)+ x\frac { \partial D(x,Q^2)}{\partial x}\right) 
\end{eqnarray}
results from a Taylor expansion of Eq.\ (\ref{Sder_full}) in the small parameter $\Delta{x}$. This equation takes into account both the gain term and the loss term in a form which is suitable for analytical investigations to understand the effects of the scattering term. This evolution equation can be solved by the standard Mellin transform \cite{Fong:1990nt}, 
$d(J,Q^2)=\int_0^1  dz\, z^{J-1}\, D(z,Q^2)$, as will be investigated in a forthcoming paper.

Here, we note that this scattering term includes the parameter $\hat q = n_g
\sigma \langle q_\perp^2\rangle$, which is well known from other Eikonal
approaches to energy loss \cite{Kovner:2003zj}. The gluon-gluon cross section of Eq.\
(\ref{eq:dsigmadt}) is of order $\sigma \approx 9\pi\alpha_s^2(Q^2)/(2 \mu_D^2)$
and the mean transverse momentum squared approximately equals the Debye mass
squared $\langle q_\perp^2 \rangle \approx \mu_D^2$  where the Debye mass is
determined by a self-consistency equation \cite{Braun:2006vd}. The
prefactor in the scattering term Eq.\ (\ref{S_der}) can then be written as
\begin{eqnarray}\label{S2}
\frac{ n_g(T) \sigma \langle q_\perp^2\rangle}{ 2 m_\text{s} Q^2} = \varepsilon  
\alpha_s^2(Q^2) \frac{T^2}{Q^2}\quad\mbox{where}\quad
\varepsilon     =\frac{36 \zeta(3)}{\pi }\sqrt{2}\frac{T}{\mu_D} 
\end{eqnarray} 
For these soft scatterings one therefore finds $\hat{S}\propto T^2/Q^2$ (recalling that $\mu_D\propto T$), which show that large $Q^2$ suppresses the effects of collisions compared to gluon splitting. This higher twist effect is, however, compensated by a large derivative in Eq.\ (\ref{S_der}) due to the steeply falling fragmentation function $D$ at $x \rightarrow 1$. 

 These results are similar to those obtained from the scattering function $\hat{K}$ in Eq.\ (\ref{K-hat}) when used in the gain and loss terms of Eq.\ \ref{Sk}. This is no accident since the general evolution equation (\ref{pt-evolution}) can be used to derive the modified DGLAP equation for evolution in the medium that was obtained in the Section \ref{sec-basics}.
To demonstrate this one starts from the scattering term in Eq.\ (\ref{Sder_full}) and 
first averages over the direction of $\vec{q}_\perp$ assuming that 
$\frac{d\sigma}{d^2\vec{q}_\perp}$ has no dependence on the azimuthal angle.
Next, by integrating out the $\vec{p}_\perp$ dependence one obtains
\begin{equation}
S(x,Q^2)
=\frac{ En_g}{Q^2}\left\{ \int_x^1 dy \int d{q}_\perp^2 
\frac{d\sigma}{d{q}_\perp^2}
yD\left(y,Q^2\right)\delta\left(y-x-\frac{{q}_\perp^2}{2m_\text{s}E}\right)
-\int d{q}_\perp^2\frac{d\sigma}{d{q}_\perp^2}
xD(x,Q^2)
\right\}
\end{equation}
where the differential cross section is different in the gain and loss terms. By performing the change of variables from ${q}_\perp$ to $z$, with the identification $-t = {q}_\perp^2=2m_\text{s} E\, y(1-z)$, 
and using the explicit form of the cross section one obtains for the gain term
\begin{eqnarray}
S_{Gain}&=&\frac{E n_g}{Q^2} \frac{9}{2} \pi \alpha_s^2(Q^2) \int_x^1 dy
\int dq_\perp^2
\frac{1}{({q}_\perp^2+\mu_D^2)^2} y D(y, Q^2) 
\delta\left(y-x-\frac{{q}_\perp^2}{2m_\text{s}E}\right)\nonumber\\
&=&\frac{E n_g}{Q^2} \frac{9}{2} \pi \alpha_s^2(Q^2) \int_x^1 dy 
\int dz\, 2 m_\text{s} y E \frac{1}{(2 m_\text{s} y E(1-z)+\mu_D^2)^2} y D(y, Q^2) \delta(yz-x)\nonumber\\
&=&\frac{n_g}{Q^2} \frac92 \pi \alpha_s^2(Q^2) \int_x^1 
\frac{dz}{z} 2 m_\text{s} \left(\frac{x}{z} E\right)^2 
\frac{1}{(2 m_\text{s} \frac{x}{z} E(1-z)+\mu_D^2)^2}  D\left(\frac{x}{z}, Q^2\right) .
\end{eqnarray}
Using Eqs.\ (\ref{P-scat}) and (\ref{K-hat}) finally gives
\begin{equation}\label{equi:gain}
S_{Gain}=\int_x^1 \frac{dz}{z} \alpha_s^2(Q^2) 
\hat{K}\left(z,\frac{xE}{z},Q^2\right) D\left(\frac{x}{z},Q^2\right)
\end{equation}
for the gain term. The loss term is treated in the same way, giving 
\begin{eqnarray}\label{equi:loss}
S_{Loss}&=&\frac{E n_g}{Q^2} \frac{9}{2} \pi \alpha_s^2(Q^2) 
\int d{q}_\perp^2\frac{1}{({q}_\perp^2+\mu_D^2)^2} x D(x, Q^2) \nonumber\\
&=&\frac{n_g}{Q^2} \frac92 \pi \alpha_s^2(Q^2) 
\int dz\, 2 m_\text{s} \left(x E\right)^2 \frac{1}{(2 m_\text{s} x E(1-z)+\mu_D^2)^2}  D\left(x, Q^2\right) \nonumber\\
&=&\int dz \alpha_s^2(Q^2) 
\hat{K}\left(z,xE,Q^2\right)D(x,Q^2) .
\end{eqnarray}
Subtracting the loss term Eq.\ (\ref{equi:loss}) from the gain term Eq.\ (\ref{equi:gain}), and including the $p_\perp$-integrated terms from the splitting in Eq. (\ref{pt-evolution}), one obtains the medium-modified DGLAP equation (\ref{Sk}).

\section {Numerical results from the evolution equations}\label{sec-numerical}
The formalism developed above applies for a gluon traversing a plasma of density $n_g(T)$ depending on temperature $T$. To solve the evolution equations for a plasma which expands and cools down with time is quite complicated. A realistic treatment of such an expanding plasma in which the traversing parton radiates and scatters requires a Monte Carlo simulation approach, which is presently being developed \cite{Zapp-MC}.
In order to obtain some numerical results to illustrate the main effects to be expected, we here simplify by considering a plasma with constant temperature $T=500$ MeV and corresponding density given by Eq.\ (\ref{eq:density}), representing the average properties of the plasma.
At this temperature $\mu_\text{D}=1.97$ GeV, obtained from $\mu_\text{D}\propto T$ in \cite{Braun:2006vd}. This will not give results that can be applied to describe heavy ion collision data, but should illustrate the main qualitative effects.

The theoretical formalism above has focussed on the parton level processes. In order to compare with observable hadrons, one has to account for hadronization. As justified above, one may use a standard parametrisation for the fragmentation function at $Q^2_0\sim 1$ GeV$^2$ obtained from data in the vacuum case. We use the KKP parametrisation \cite{Kniehl:2000fe} for the hadronisation of gluons into charged pions 
\begin{equation}\label{KKP}
D_\text{g}^\pi(x,Q_0^2)=3.73\,x^{-0.742}\,(1-x)^{2.33}
\end{equation}
at the scale $Q_0^2 = 2 \mbox{GeV}^2$ in our evolution equations. 

\begin{figure}[ht]
\centering
{\footnotesize
\psfrag{$x$}{\footnotesize $x$}
\psfrag{ylabel}{\footnotesize $D_\text{g}^\pi (x,Q^2)$}
\psfrag{label1}{\footnotesize $Q_0^2=\unit[2]{GeV^2}$}
\psfrag{label2}{\footnotesize $Q^2=(\unit[10]{GeV})^2$}
\psfrag{label3}{\footnotesize $Q^2=(\unit[100]{GeV})^2$}
\psfrag{label4}{\footnotesize $Q^2=(\unit[100]{GeV})^2$ vacuum}
\psfrag{label5}{\footnotesize $Q^2=(\unit[100]{GeV})^2$ medium}
\psfrag{$10^{-7}$}{\footnotesize $10^{-7}$}
\psfrag{$10^{-6}$}{\footnotesize $10^{-6}$}
\psfrag{$10^{-5}$}{\footnotesize $10^{-5}$}
\psfrag{$10^{-4}$}{\footnotesize $10^{-4}$}
\psfrag{$10^{-3}$}{\footnotesize $10^{-3}$}
\psfrag{$10^{-2}$}{\footnotesize $10^{-2}$}
\psfrag{$10^{-1}$}{\footnotesize $10^{-1}$}
\psfrag{$10^0$}{\footnotesize $10^0$}
\psfrag{$10^1$}{\footnotesize $10^1$}
\psfrag{$10^2$}{\footnotesize $10^2$}
\psfrag{$10^3$}{\footnotesize $10^3$}
\psfrag{ 0}{\footnotesize 0}
\psfrag{ 0.1}{\footnotesize 0.1}
\psfrag{ 0.2}{\footnotesize 0.2}
\psfrag{ 0.3}{\footnotesize 0.3}
\psfrag{ 0.4}{\footnotesize 0.4}
\psfrag{ 0.5}{\footnotesize 0.5}
\psfrag{ 0.6}{\footnotesize 0.6}
\psfrag{ 0.7}{\footnotesize 0.7}
\psfrag{ 0.8}{\footnotesize 0.8}
\psfrag{ 0.9}{\footnotesize 0.9}
\psfrag{ 1}{\footnotesize 1}
\includegraphics{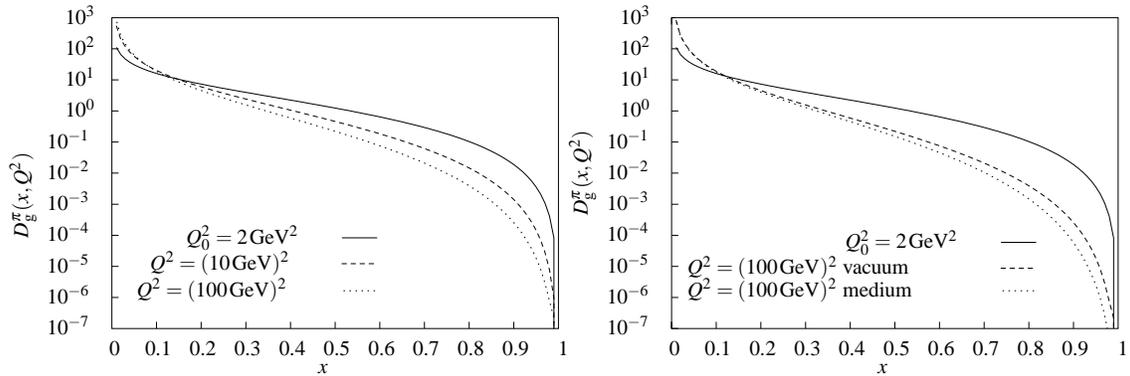}}
\caption{\label{fig_vff} 
The fragmentation function of gluons into charged pions obtained when solving Eq.\ (\ref{Sk}) for different initial virtualities $Q^2$ and with the boundary condition at the perturbative QCD cut-off $Q^2_0=2$ GeV$^2$ given by the KKP parameterization in Eq.\ (\ref{KKP}) for hadronisation. 
Left without scatterings, i.e.\ vacuum case with conventional DGLAP $Q^2$ evolution. Right with medium-modified evolution including scatterings of gluons, with initial energy $E=Q_{max}$, in a gluon plasma of fixed temperature $T=500\mbox{MeV}$ compared to the vacuum case.
}
\end{figure}

The resulting gluon-to-pion fragmentation functions are shown in Fig.  \ref{fig_vff} for both the normal vacuum case and for the case of a plasma using our medium-modified DGLAP equation. The overall behaviour is similar. Already in the vacuum case, the evolution caused by a higher $Q^2$ depletes the high $x$ region and enhances the low $x$ part, since the parton cascade means that energy is shared among more and more partons. The scattering term acts in a similar manner and ``transports'' partons to a lower momentum fraction.
Our scattering formalisms account for the energy loss of the gluon traversing the medium, but not the fact that the struck scattering centers obtain energy that may contribute to the jet energy. Such struck partons are expected to interact further in the plasma and contribute to the underlying activity in the event, including possible minijets, but they will not produce leading hadrons in the high-$p_\perp$ jets which is the focus of this study.

\begin{figure}[ht]
\centering
{\small
\psfrag{xlabel}{$x$}
\psfrag{ylabel}{$D_\text{g,med}^\pi (x,Q^2)/D_\text{g,vac}^\pi (x,Q^2)$}
\psfrag{label1}{$Q^2 = (\unit[10]{GeV})^2$}
\psfrag{label2}{$Q^2 = (\unit[100]{GeV})^2$}
\psfrag{ 0}{0}
\psfrag{ 0.1}{0.1}
\psfrag{ 0.2}{0.2}
\psfrag{ 0.3}{0.3}
\psfrag{ 0.4}{0.4}
\psfrag{ 0.5}{0.5}
\psfrag{ 0.6}{0.6}
\psfrag{ 0.7}{0.7}
\psfrag{ 0.8}{0.8}
\psfrag{ 0.9}{0.9}
\psfrag{ 1}{1}
\includegraphics{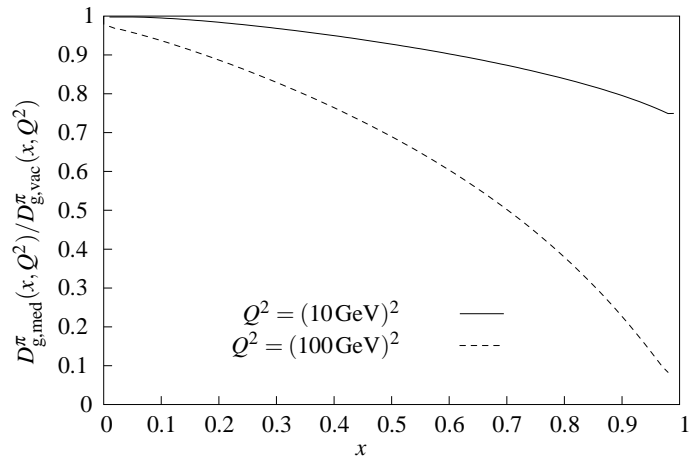}}
\caption{\label{fig_ffratios}
Ratios of fragmentation functions in medium and vacuum, 
$D_{g,med}^\pi (x,Q^2)/D_{g,vac}^\pi (x,Q^2)$, of gluons into charged pions for initial gluon energies $E=Q_{max}=10$ and 100 GeV (cf.\ Fig. \ref{fig_vff}). The medium fragmentation function is calculated using the modified DGLAP evolution, Eq.\ (\ref{Sk}), including scatterings in a gluonic plasma of $T=500$ MeV.
}
\end{figure}

\begin{figure}[ht]
\centering
\includegraphics[width=0.6\textwidth]{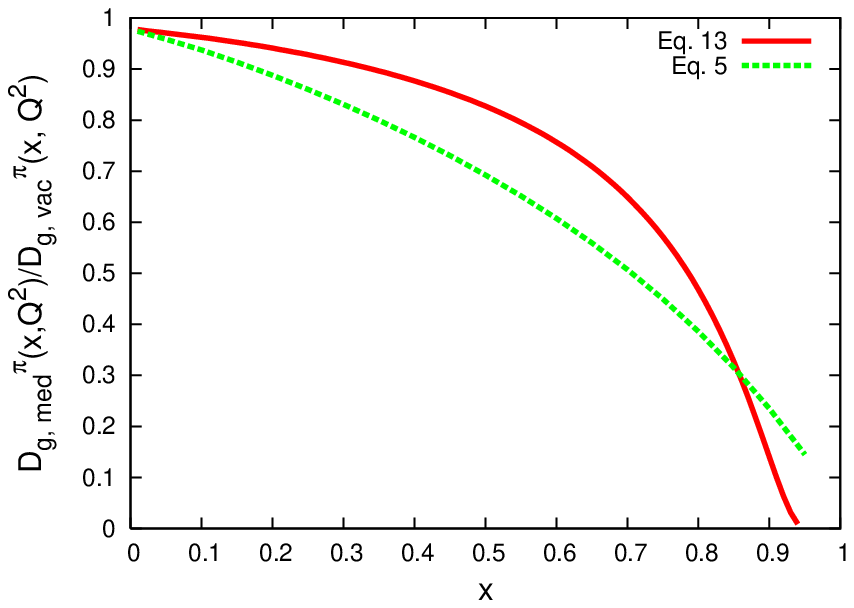}
\caption{\label{comp} 
Ratios of fragmentation functions in medium and vacuum, 
$D_{g,med}^\pi (x,Q^2)/D_{g,vac}^\pi (x,Q^2)$, of gluons into charged pions at $Q^2=(100\mbox{ GeV})^2$ where the scatterings in the medium (gluon plasma with $T=500\mbox{ MeV}$) are included via our medium-modified evolution Eq.\ (\ref{Sk}) (dashed curve) and Eq.\ (\ref{S_der}) (full curve).
}
\end{figure}

A more detailed comparison of the fragmentation function in the medium and in
the vacuum is given in Fig.~\ref{fig_ffratios} by their ratio, which shows a
moderate suppression of leading particles in a \unit[10]{GeV} jet
and a strong one in the case of a \unit[100]{GeV} jet.  The stronger
quenching effect in the latter case depends not only on the longer evolution
path in $\ln Q^2$, but also on the energy dependence of the
scattering, as illustrated in Eq.\ (\ref{Sk4}). Thus, collisional energy
loss is certainly non-negligible, even using a perturbative cross section, and 
can become quite large, as also indicated in an earlier study \cite{Zapp:2005kt} 
based on a phenomenological model for soft colour interactions. The
total energy loss will be even larger when additional medium-induced radiation
is taken into account \cite{wang, gyulassy, wiedemann}.

This ratio of fragmentation functions in medium and vacuum corresponds to the
nuclear modification factor $R_\text{AA}$ at a fixed jet energy. The
$R_\text{AA}$ measured in heavy ion collisions at RHIC \cite{whitepapers} is for leading hadrons with $p_\perp$ up to $\sim 20$ GeV, and thereby integrates over jet (parton) energy. 
For a direct comparison with RHIC data, one therefore needs to fold this kind of fragmentation functions for both gluons and quarks with the initial momentum distribution of gluons and quarks from the hard scattering. A detailed comparison to data also requires a proper treatment of the finite-size expanding plasma. 
Taking this into account is beyond the scope of this paper, which does not aim at giving a precise prediction to be compared with experimental data but rather to develop a formalism and show its basic physical effects. Nevertheless, our approximation with a stationary plasma in the numerical solution of the evolution equations is reasonable to illustrate the magnitude of jet quenching effect. To see this one has to consider the time needed for a parton shower to evolve down to $Q_0$. This depends, however, on the topology of the shower. In the evolution equation all possible radiation patterns are weighted with their probabilities and summed, making it impossible to keep precise track of the time during the evolution. Based on Eq.~(\ref{dtau}), the mean time for evolution can be estimated as $E/Q_0^2-1/E$, which means that a \unit[100]{GeV} jet developes over \unit[10]{fm} to reach $Q_0=\unit[\sqrt{2}]{GeV}$. A finite size medium can then be included by defining two cut-off scales, $Q_1$ at which scatterings stop and $Q_0$ at which splittings stop. The suppression factor is then found to vary roughly linearly with the path length $L\simeq E/Q_1^2-1/E$ in the plasma. For instance, a medium of length $L=\unit[7]{fm}$ (corresponding to
$Q_1^2\approx 3$ GeV$^2$) instead of $L=\infty$ results in a
 suppression factor of 0.57 instead of 0.38 for leading particles at
 $x=0.8$ in a \unit[100]{GeV} jet. The $x$-averaged suppression factor is  0.75
for $L=\unit[7]{fm}$ instead 
of 0.64 for $L=\infty$, \ie\ overall only an approximately $15$\% difference. We
note that these differences are of similar magnitude as the difference
between our two forms of the scattering term shown in
Fig.~\ref{comp} below.

Furthermore, the recoiling scattering centre would also have to be taken into account.
These complications can best be studied using the Monte Carlo framework under development \cite{Zapp-MC}.
Already at this stage one can, however, note that there is an overall general qualitative agreement, but the effect in data is much stronger than the suppression of a 10 GeV jet shown in Fig.~\ref{fig_ffratios}. This indicates the need for adding medium-induced radiation and raises the question whether the medium affects also the non-perturbative hadronisation.

The strongly enhanced effect of collisional energy loss at higher jet energies 
provides interesting prospects for heavy ion collisions at LHC, where individual
hadrons and jets with $p_\perp$ up to $\sim 100$ GeV will become available.
The curve for $Q=100$ GeV in Fig.~\ref{fig_ffratios} should not be taken as a
precise prediction, in view of the uncertainties mentioned. It does, however,
indicate a substantially increased effect of energy loss due to collisions. 
This is due to the enlarged phase space of momentum transfer 
squared $|t|$, a longer evolution path in 
$\ln Q^2$, the longer parton lifetime and, 
as discussed in connection with 
Eq.\ (\ref{K-hat}), the enhanced scattering function $\hat{K}$ for large 
parton energy which can give multiple scatterings that add up small 
energy losses to a substantial effect.

The magnitude of this effect is checked in Fig. \ref{comp} using the 
alternative form of the scattering function in Eq.\ (\ref{S_der}). 
It is reassuring to see that the behavior as a function of $x$ is 
overall the same, although somewhat different in the details. 
The difference between the two curves stems essentially from the 
Taylor expansion in Eq.\ (\ref{S_der}) as well as the different 
allowed ranges for the momentum transfer squared $|t|$. 
As expected from this equation, 
the effect of the scattering becomes large at $x\to 1$.

Remembering that the quark-gluon plasma is here approximated by a stationary 
medium of average temperature $T=500$ MeV, one may worry that taking into 
account the expansion and cool-down of the plasma might lead to a smaller 
collisional energy-loss effect. This need not be the case, since the multiple 
scattering that can occur even in a short, early time interval when the plasma 
is denser and hotter than average, has an opposite effect. 

Jet multiplicities can be simply obtained by integrating the fragmentation
function $D(x,Q^2)$ over $x$, but we do not consider it meaningful to give
numerical results on this without including the suppression of soft gluon
radiation. This can be done, as in the vacuum case, by replacing the gluon
distribution $D(z/u,Q^2)$ in the DGLAP evolution term by $D(z/u,u^2 Q^2)$ with a reduced virtuality scale \cite{Fong:1990nt}. 
This will be studied in a forthcoming publication.

\section{Concluding discussion}\label{sec-conclusions}
To summarize, we have developed a formalism that combines the conventional parton cascade with parton scattering in a plasma of varying density and temperature. In order to give first numerical results, we simplified by assuming a plasma of constant temperature and density. This clearly demonstrated that collisional energy loss contributes substantially to jet quenching. The formalism developed can be included in a complete framework, where a proper expansion of the quark-gluon plasma as well as the momentum distributions of hard-scattered partons are included, e.g.\ in a Monte Carlo simulation program \cite{Zapp-MC}. This would allow a detailed comparison to RHIC data on jet quenching. Already at this stage, it seems clear that although the overall behaviour is in agreement with the observed one, the magnitude is smaller than observed at RHIC. This leaves room for the expected additional medium-induced radiation \cite{wang, gyulassy, wiedemann}.

For RHIC conditions, with initial parton energy $10\mbox{ GeV}<E<20\mbox{ GeV}$, the jet quenching effect obtained here does not vary strongly with this energy but depends essentially only on the momentum fraction. This gives a quenching effect which is essentially independent of the transverse momentum of the hadrons and thereby a nuclear modification factor $R_{AA}$ which is flat over a substantial $p_\perp$-interval, just as observed in the RHIC data. 

For heavy ion collisions at LHC we find a substantially increased jet quenching 
effect at higher jet energies due to the increased range of allowed
momentum transfers, an increased evolution path in $\ln{Q^2}$, 
 the longer lifetime of the parton as well as of the plasma, 
and enhanced scattering probability. 
 Multiple scatterings may then add up many relatively small energy losses to a 
substantial effect.  Thus, 
we expect,  e.g., the nuclear modification factor $R_{AA}$ to be further 
reduced at the higher $p_\perp$ available at LHC.

A comparison of jet profiles of heavy quarks with gluon jets may help to fix the form of the plasma scattering cross section more precisely. Since the momentum transfers extend to low momenta $\propto T$, a purely perturbative calculation of the cross section may be questionable and an improved treatment needed  \cite{Braun:2006vd}. Especially one has to study further how the temperature scale $T$ can modify the leading $T^2/Q^2$ behaviour.

A recent calculation of jet evolution in the modified leading log approximation \cite{Borghini:2005mp} has produced differential multiplicity distributions. The advantage of our calculation is that it takes into account the scattering term explicitly and therefore gives results which depend on the plasma density and temperature. Taking transverse momentum into account explicitly as an extra variable, the equation could in principle be used to investigate the $\vec{p}_\perp$ broadening of the parton in the medium, which is highly relevant for the size of the jet cone in nucleus-nucleus collisions and will be studied further.

Another issue to be studied is whether the plasma affects the non-perturbative hadronisation processes and thereby changes the fragmentation function. This may depend on whether partons are first forming preconfined states \cite{Accardi:2002tv,Kopeliovich:2007dt}, as is theoretically supported e.g.\ in the large $N_c$ approximation, or directly form hadrons (including resonances). Intermediate prehadrons below the scale $Q_0$ 
may suffer more scatterings than final hadrons and thereby modify the hadronic spectra of heavy ion collisions. The non-perturbative phase after evolution down to $Q=\sqrt{2}$ GeV should be relatively more important at the lower energy scale of RHIC. At LHC, the larger energy implies that the formation of prehadrons and hadrons occurs later in lab frame, in particular for leading hadrons with a large Lorentz $\gamma$ factor. We therefore expect that medium-modified hadronisation effects should be less and thereby our theoretical approach closer to reality.

To conclude, we have developed a formalism to account for energy losses due to scattering in a plasma. With the indicated further developments this should facilitate systematic comparisons with the observed jet quenching in heavy ion collisions, which should contribute to a demonstration of the existence of the quark-gluon plasma and an exploration of its properties.

\begin{acknowledgments}
  Useful conversations with N.\ Borghini and U.\ Wiedemann are gratefully acknowledged. This work was supported by DFG PI 182/3-1, BMBF 06HD196 CERN ALICE
  and within the framework of the Excellence Initiative by the German
  Research Foundation (DFG) 
  through the Heidelberg Graduate School of Fundamental Physics 
  (grant number GSC 129/1). Also financial support from the Swedish Research
  Council and The Swedish Foundation for International Cooperation in Research and Higher Education (STINT) is gratefully acknowledged.
\end{acknowledgments}

\end{document}